\documentclass[11pt]{article}

\usepackage[iso-8859-7]{inputenc}
\usepackage{epsfig}
\usepackage{graphicx}
\usepackage{epstopdf}
\usepackage{amsfonts}
\usepackage{amssymb}
\usepackage{amsthm}
\usepackage{amsmath}
\usepackage{multirow}

\usepackage{subcaption}

\newcommand{\newc}{\newcommand}
\newc{\ra}{\rightarrow}
\newc{\lra}{\leftrightarrow}
\newc{\be}{\begin{equation}}
\newc{\ee}{\end{equation}}
\newc{\bs}{\begin{split}}
	\newc{\es}{\end{split}}
\newc{\ba}{\begin{eqnarray}}
\newc{\ea}{\end{eqnarray}}
\newc{\ov}{\overline}
\newc{\pa}{\partial}
\newc{\D}{\Delta}

\newc{\nn}{\nonumber}

\begin{document}
	\begin{titlepage}
	
	\vspace*{0.7cm}

	\begin{center}
		{\Large {\bf Vectorlike Particles, $Z'$ and Yukawa Unification in  F-theory inspired $E_6$  }}
		\\[12mm]
	Athanasios Karozas$^{a}$ \footnote{E-mail: \texttt{akarozas@cc.uoi.gr}}, George K. Leontaris$^{a}$
		\footnote{E-mail: \texttt{leonta@uoi.gr}} and
		Qaisar Shafi$^{c}$
		\footnote{E-mail: \texttt{shafi@bartol.udel.edu}}
		\\[-2mm]
		
	\end{center}
	\vspace*{0.50cm}
	\centerline{$^{a}$ \it
		Physics Department, Theory Division, University of Ioannina,}
	\centerline{\it
		GR-45110 Ioannina, Greece }
	\vspace*{0.2cm}
	\centerline{$^{c}$ \it
		Bartol Research Institute, Department of Physics and Astronomy, University of Delaware,}
	\centerline{\it
		DE 19716,  Newark, USA}
	\vspace*{1.20cm}
	
	\begin{abstract}
		\noindent
		
 We explore the low energy implications of an F-theory inspired $E_6$ model whose breaking yields, in addition  to the 
 MSSM gauge symmetry, a $Z'$ gauge boson associated with a $U(1)$ symmetry broken at the TeV scale. The zero mode spectrum 
 of the effective low energy theory is derived from the decomposition of the $27$ and $\overline{27}$  representations of 
 $E_6$ and we parametrise their multiplicities in terms of a  minimum  number of flux parameters.  We perform a two-loop 
 renormalisation group analysis of the gauge and Yukawa couplings of the  effective theory model and  estimate lower bounds
  on the new vectorlike particles predicted in the model. We compute the third generation Yukawa couplings in an F-theory  
 context assuming an $E_8$ point of enhancement and express our results in terms of the local flux densities associated with 
  the gauge symmetry breaking. We find that their values are compatible with the ones  computed by the renormalisation group
 equations, and we identify points in the  parameter space of the flux densities  where the $t-b-\tau$  Yukawa couplings unify.

	\end{abstract}

\end{titlepage}

\section{Introduction}

\setcounter{footnote}{0}

The existence of a neutral  gauge boson  $Z'$  associated with a new  $U(1)$ gauge symmetry spontaneously broken at  a few TeV  
is an interesting possibility. It is well-motivated both experimentally as well as theoretically, and  its implications have been 
extensively discussed in the literature~\cite{Cvetic:1997ky,Langacker:2008yv,Cvetic:2011iq}.  
The experimental bound on the mass of a $Z'$ boson  decaying  only to ordinary  quarks and leptons  with 
couplings comparable to the Standard  Model (SM) $Z$ boson, is about 
 $ 3$~TeV~\cite{Khachatryan:2017wny,Aaboud:2017yvp,ATLAS:2017wce}. Theoretically,  several extensions of the Standard Model and their 
 supersymmetric versions,  predict the  existence  of additional  $U(1)$  symmetries.  In the context of Grand Unified Theories (GUTs)  
 these are  embedded in  gauge groups larger than $SU(5)$  since the latter  contains only  the SM  gauge group. 
  
 One of the most interesting unified  groups containing additional abelian factors of phenomenological interest is the exceptional group 
 $E_6$~\cite{Gursey:1975ki,Achiman:1978vg,Shafi:1978gg}.  This has been extensively studied  as a field theory unified model as well as 
in a string background. It emerges naturally in many string compactifications and, in particular, in an F-theory framework~\cite{Beasley:2008dc},
where several interesting features have been 
discussed~~\cite{Callaghan:2013kaa,Callaghan:2011jj,Chen:2010tg,Cvetic:2012ts,Leontaris:2016wsy}.
 Under the breaking  pattern $E_6\supset SU(5)$,
   two  abelian factors appear, usually dubbed   $U(1)_{\chi}$ and $U(1)_{\psi}$. In general, after the spontaneous symmetry breaking of $E_6$, some 
   linear  combination   of these $U(1)$'s  may survive at low energies~\cite{Langacker:1998tc}.  The corresponding neutral gauge boson receives 
   mass at the TeV scale and may be found at LHC or its upgrates.

 In this work we examine the implications  of a TeV scale neutral gauge  boson  corresponding to various possible combinations of $U(1)_{\psi}$ and $U(1)_{\chi}$.  In addition,  motivated by string and in particular F-theory effective models,  we consider the existence of additional vectorlike 
 fields  and  neutral singlets at the TeV scale.  We assume that the initial $E_6$ symmetry is broken by background fluxes which leave  only one 
 linear $U(1)$ combination  unbroken, commutant with $SU(5)$. In the present work the zero mode spectrum of the effective theory is derived from 
 the decomposition of  the $27$ and  $\overline{27}$ representations of $E_6$, and, we parametrise their multiplicities in terms of a  minimum  
 number of (integer) flux parameters.  In addition, since  the flux-breaking mechanism  splits the $E_6$ representations into incomplete 
  multiplets~\cite{Callaghan:2013kaa,Callaghan:2011jj,Chen:2010tg,Cvetic:2012ts,Leontaris:2016wsy},  one may  choose appropriately the flux 
  parameters in order to retain only the desired components from the 27 and $\overline{27}$ representations. 

We also perform a two-loop renormalisation group equations (RGE) analysis of the gauge and Yukawa couplings of the  effective theory model  
for different choices 
of linear combinations of the $U(1)$ symmetries.  Implementing the idea of incomplete $E_6$ representations  motivated by F-theory considerations, 
we make use of zero mode spectra  obtained from  truncated $E_6$ representations. We use known mathematical packages~\cite{Staub:2013tta}, to derive 
and solve numerically  the  RGE's in the presence of additional matter such as vectorlike triplets, doublets and singlet fields with masses
 down to the TeV scale.  Furthermore, we investigate possible gauge and Yukawa coupling unification  by considering four different cases with respect 
 to the unbroken  $U(1)$ combination after breaking $E_6$ down to the SM.  Finally, we perform an  F-theory  computation of the Yukawa couplings at 
 the GUT scale and express them in terms of the various local flux parameters
 associated with the symmetry breaking.

	\section{ $E_6$ GUT in an F-theory perspective }
	
We start with a short description of the $E_6$ GUT breaking  and the massless spectrum. The $U(1)$ symmetries we are interested in
	appear under the breaking pattern 
 \begin{align}
 E_{6}&\rightarrow{SO(10)\times{U(1)_{\psi}}}\rightarrow{SU(5)\times{U(1)_{\psi}}\times{U(1)_{\chi}}}.\label{E62SU5}
\end{align}	
In  an effective $E_6$ model with an F-theory origin,  matter fields, in general, arise from ${ 27}, \overline{ 27}$ and ${78}$ representations.
In the present work we restrict to the case where the three families, the Higgses and other possible matter fields emerge from the decomposition 
of the ${ 27}(\in E_6)$ under $SO(10)\times U(1)_{\psi}$,
\be 
{ 27}\to {16}_1+{10}_{-2}+ { 1}_{4} .\label{27dec}
\ee 
The decompositions of the $SO(10)$ multiplets in (\ref{27dec}) under the breaking of $SO(10)$ to $SU(5)$ are as follows
\ba 
 {16}_1\to10_{(1,-1)}+\bar 5_{(1,3)}+1_{(1,-5)},\;
 {10}_{-2}\to5_{(-2,2)}+\bar 5_{(-2,-2)},\;
 { 1}_{4}\to (1,1)_{(4,0)},
 \ea 
 where the two indices respectively refer to the charges under the two abelian factors $U(1)_{\psi} \times U(1)_{\chi}$.
 
The  fermion families  are accommodated in three  $16$-plets of $ SO(10)$. The ordinary quark triplets, the right-handed  
electron and lepton doublets comprise the  $10_{(1,-1)}$ and $ \bar 5_{(1,3)}$  of $SU(5)$, and  in the standard description, the singlet  
$1_{(1,-5)}$ is identified with the right-handed  neutrino.  There are also  vectorlike multiplets  $5_{(-2,2)}+\bar 5_{(-2,-2)}$ 
and  $SO(10)$ singlets with charges ${(4,0)}$.   The normalised charges $\tilde Q_a= N_a Q_a$
are defined so that	${\rm Tr}\,\tilde Q^2_{a}= 3$, and therefore $ N_\psi=\frac{1}{2\sqrt{6}}$ and $ N_{\chi}=\frac{1}{2\sqrt{10}}$.

With the spontaneous breaking of $ U(1)_{\psi}$ and  $U(1)_{\chi}$,  the corresponding neutral gauge bosons 
 receive masses of the order of their breaking scale.  Depending on the details of the particular model, the breaking scale of 
 these $U(1)$'s can be  anywhere between 
$M_{GUT}$ and a few TeV, with the latter determined  by LHC. New Physics phenomena can be anticipated in the
TeV range and  possible deviations of the SM predictions are associated with the existence of a new neutral gauge 
boson in this range. In the present model, a $Z'$ boson that may appear at low energies could be any linear combination 
of the  form $Z'=Z_{\chi}\cos\phi+Z_{\psi}\sin\phi$.  The corresponding $U(1)$ charge  is  defined by
\be
Q=\tilde  Q_{\chi}\cos\phi + \tilde  Q_{\psi}\sin\phi .\label{Qdef}
\ee

Several values of the mixing angle $\phi$ lead to  models consistent with the data. The following models are of our primary interest in this work. 

$\bullet$ 
N-model  \cite{Ibanez:1986si,Ma:1995xk,King:2005jy}: We assign the right-handed  neutrinos in $1_{(1,-5)}$, and  require $Q_{\nu}=0$.
Then, from (\ref{Qdef}), we fix $\tan\phi=\sqrt{15}$ and as a result,
\be 
 Q_N= \frac{1}{4}\sqrt{\frac 58}\left( Q_{\psi}+\frac 15 Q_{\chi}\right).\label{QNdef}
\ee 

$\bullet$ 
$\eta$-model:  In this case the $U(1)_{\eta}$ charge formula takes the form 
\be 
 Q_{\eta}= -\frac{1}{8}\sqrt{\frac 53}\left( Q_{\psi}-\frac 35 Q_{\chi}\right),\label{Qetadef}
\ee 

\noindent which arises as a consequence of breaking $E_6$ directly to a rank-5 group \cite{Witten:1985xc}.

$\bullet$   $\chi$-model where $\phi=0$, and  $\psi$-model where $\phi=\pi/2$.

The phenomenological implications of these models have recently been discussed in~\cite{ Athron:2015vxg, Athron:2016qqb,Hicyilmaz:2016kty,Hebbar:2016gab}, while an analysis with a general mixing angle, $\phi$, is presented in~\cite{ Belanger:2017vpq, Araz:2017qcs,Hicyilmaz:2017nzo}.  The $(\psi, N, \eta)$-charges of the $SU(5)$ representations are shown in Table~\ref{charges}. Details for the $\chi$-model are presented 
separately  in Table \ref{XXX} since we use a different GUT origin for the SM spectrum.  (Notice that $Q_{\chi}=-Q_{N}$ and, as a result, the 
RGE analysis presented  in the next sections is  the same.)

Having described the basic features of the models, we proceed now to the
derivation of the spectrum from F-theory perspective. 
\begin{center}
\begin{table}[t]
\centering
\begin{tabular}{|ccc|ccc|c|c|}
\hline
$E_6$ & $SO(10)$ & $SU(5)$  &$\sqrt{24}Q_{\psi}$&$\sqrt{10}Q_{N}$ & $\sqrt{15}Q_{\eta}$ &SM \\
\hline
$27$ & $16$ & $\overline{5}_M$ & $\hphantom{+} 3$  &$\hphantom{+} 1$ & $\hphantom{+}\frac 12$& $d^{c}, L$\\
$27$ & $16$ & $10_M$  &$\hphantom{+} 1$& $\hphantom{+} \frac 12 $&$-1$ &$Q, u^c, e^c$\\
$27$ & $16$ & $1_{\nu}$ &$\hphantom{+} 1$&$\hphantom{+} 0$ &$-\frac 52$ &  $\nu^c$\\
$27$ & $10$ & $5_H$  & $-2 $&$-1$ &$ \hphantom{+}2$& $D, H_u$\\
$27$ & $10$ & $\overline{5}_{\bar H}$ &$ -2$ &$-\frac 32 $&$\hphantom{+} \frac 12$ & $\overline{D}, H_d$\\
$27$ & $1$ & $\hphantom{+} 1$  & $\hphantom{+} 4 $&$\hphantom{+}\frac 52$ &$-\frac 52$& $S$\\
\hline
\end{tabular}
\caption{\small  $27$ of $E_6$ and its $SO(10)$ and $SU(5)$ decompositions and $Q_{\psi, N, \eta }$ charges.  }
\label{charges}
\end{table}
\end{center}
\begin{center}
\begin{table}[t]
\centering
\begin{tabular}{|c|c|c|c|c|c|}
\hline
$E_6$ & $SO(10)$ & $SU(5)$ & $\sqrt{10}Q_{\chi}$ & SM particle content\\
$27$ & $10$ & $\overline{5}_M$ & $-1$  & $d^{c}, L$\\
$27$ & $16$ & $10_M$  & $-\frac{1}{2}$ &$Q, u^c, e^c$\\
$27$ & $1$ & $1_{\nu}$ & $\hphantom{+} 0 $ &  $\nu^c$\\
$27$ & $10$ & $5_H$  & $\hphantom{+}1$  & $D, H_u$\\
$27$ & $16$ & $\overline{5}_{\bar H}$ & $\hphantom{+}\frac{3}{2}$ & $\overline{D}, H_d$\\
$27$ & $16$ & $1$  & $-\frac{5}{2}$  & $S$\\
\hline
\end{tabular}
\caption{\small  $27$ of $E_6$ and its $SO(10)$ and $SU(5)$ decompositions and $Q_{\chi}$ charges.  }
\label{XXX}
\end{table}
\end{center}

\section{F-theory motivated $E_6$ spectrum }
 
 In  F-theory,  the gauge symmetry is a subgroup of  $E_8$, the latter being associated with the highest singularity 
 of the elliptically fibred internal space. We assume that the internal manifold is equipped with 
 a divisor possessing an  $E_6$ singularity, thus  
\be 
E_8\supset E_6\times SU(3)_{\perp}. \label{E8E6U3}
\ee 
  The representations of the effective theory model, arise from the decomposition of  $ E_8$
 adjoint
 \[248\ra (78,1)+(1,8)+(27,3)+(\ov{27},\bar 3).\]
 In the above decomposition, we are interested in the zero modes $(27,3)+(\ov{27},\bar 3)$ 
 lying  on the Riemann surfaces formed on the intersections of seven branes with the $E_6$ divisor. 
 Restricting to  specific cases of GUT surfaces, such as del Pezzo or Hitzebruch, one can
 determine  the  chirality $27-\ov{27}$  in terms of a topological index, the Euler characteristic.  
We  assume the breaking of $E_6$ to the standard $SO(10)$ model by a non-trivial flux along $U(1)_{\psi}$.
 Since $E_8\supset E_6\times SU(3)_{\perp}$, the $27$'s reside on  three
  matter curves corresponding to the Cartan roots $t_i$ of $SU(3)_{\perp}$, with $t_1+t_2+t_3=0$, and  
  this implies that the only invariant Yukawa coupling is $27_{t_1}27_{t_2}27_{t_3}$. We  choose to accommodate
  the Higgs fields in $27_{t_3}=27_H$ and therefore the chiral families are on the $t_1, t_2$ curves. However, 
 in order to achieve a rank-one mass matrix  and 
 obtain a tree-level Yukawa coupling for the third generation, two matter curves have to be 
 identified, and this can be achieved   under the action of a $Z_2$ monodromy such that $t_1=t_2$.
 Furthermore, choosing appropriately the restrictions of the flux parameters 
 on the matter curves,   we can arrange things so that the  spectrum contains three families in $16(\to 10+\bar 5+1)$,
 and three Higgs pairs in $10(\to 5+\bar 5)$  and several neutral  singlets~\cite{Leontaris:2016wsy}.

Indeed, if we generally assume  that the topological characteristics of the chosen manifold  allow $M$ copies of $27_{t_1}$ and 
$M_H$ copies of $27_{t_3}$ representations on the corresponding matter curves, turning on a suitable $U(1)_{\psi}$-flux of $n$ 
and $ m$ units respectively, we get the splitting shown in  Table~\ref{fluxmultiplicities}.

\begin{center}
\begin{table}[h]
\centering
\begin{tabular}{|l|l|}
 \hline
{\rm  Matter}&{\rm  Higgs}\\
 \hline
$ 27_{t_1}/ \overline{27}_{-t_1}\begin{array}{|ll}SO(10)\times U(1)_{\psi}&\#\\
  \#(16_1- \ov{16}_{-1})&M\\
 \#(10_{-2}- \ov{10}_{2})&M+n\\
  \#(1_{4}- \ov{1}_{-4})&M-n\end{array}$
 & $27_{t_3}/ \overline{27}_{-t_3}
 \begin{array}{|ll}SO(10)\times U(1)_{\psi}&\#\\ 
 \#(16_1^H- \ov{16}_{-1}^H)&M_H\\
  \#(10_{-2}^H- \ov{10}_{2}^H)&M_H+m\\
    \#(1_{4}^H- \ov{1}_{-4}^H)&M_H-m\end{array}$
 \\
 \hline
\end{tabular}
\caption{\small Splitting of $27_{t_{1}}$ ($\overline{27}_{-t_{1}}$) and  $27_{t_{3}}$ ($\overline{27}_{-t_{3}}$) representations by turning on a suitable $U(1)_{\psi}$-flux of $n$ and $m$ units respectively.}
\label{fluxmultiplicities}
\end{table}
\end{center}

  
The spectrum also includes  singlets which descend from the $SU(3)_{\perp}$ adjoint decomposition,
designated as 
 \[  1_{t_i-t_j}\equiv \theta_{ij} , \; i,j=1,2,3 .\] 
  
As an illustration, we present two cases with minimal spectra of $E_6$ motivated models for two specific 
choices of the fluxes.
 
\noindent $1.$ An economical model emerges if we choose 
\be\label{casef1}
 M=3, \,M_H=0,\, n=-m=-3 .
 \ee
\noindent $2.$  An alternative possibility may arise if we choose
\be\label{casef2} 
 M=3,\, M_H=0, \, n=-m=-4 .
 \ee
  \begin{center}
\begin{table}[t]
\centering
\begin{tabular}{|l|l|}
 \hline
  	{\rm  Matter}&{\rm  Higgs}\\
  	\hline
 $ 	27_{t_1}/ \overline{27}_{-t_1}\begin{array}{|lrr}SO(10)\times U(1)_{\psi}&\#1&\#2\\ 16_{1}&3&3\\\overline{10}_{2}&0&1\\1_{4}&6&7\end{array}$
  	& $27_{t_3}/ \overline{27}_{-t_3}\begin{array}{|lrr}SO(10\times U(1)_{\psi}&\#1&\#2\\16_{1}&0&0\\10_{-2}&3&4\\1_{-4}&3&4\end{array}$
  	\\
  	\hline
 \end{tabular}
\caption{\small  Two different cases of $E_6$ motivated models. The two cases labelled here as $\#1$ and $\#2$  correspond to the choice of flux parameters in equations (\ref{casef1}) and (\ref{casef2}) respectively. }
\label{fcases}
\end{table}
\end{center}
  Both cases are shown in Table \ref{fcases}. The models differ with respect to the number of $10$-plets and singlets;
however the number of 16-plets is always three. 
 In the first  choice, all $10$-plets reside on  $27_{t_3}$ Higgs curve, while in the second case there is an additional pair descending from  $\overline{27}_{-t_1}+27_{t_3}$.

 Similarly, further symmetry breaking of the $SO(10)\rightarrow{SU(5)\times{U(1)_{\chi}}}$ will be achieved by turning on suitable $U(1)_{\chi}$ fluxes~\cite{Leontaris:2016wsy}. 
Thus,  for the two $16$'s,  in general, we have 
\begin{equation}
{16}_{1}=
\left\{\begin{array}{ll}{\rm
Rep}&{\rm flux \, units }\\
\ 10_{-1}&\;3\\
\ \bar{5}_{3}&\;3+n_{\chi}\\
\ 1_{-5}& \;3-n_{\chi}\\
\end{array}\right.\quad{,}\quad  {16}_{1}^{H}=
\left\{\begin{array}{ll}{\rm
Rep}&{\rm flux \, units }\\
\ 10_{-1}&\;0\\
\ \bar{5}_{3}&\;0+m_{\chi}\\
\ 1_{-5}& \;0-m_{\chi}\\
\end{array}\right.\,,
\end{equation}
where the integers $n_{\chi}, m_{\chi}$ represent the $U(1)_{\chi}$ fluxes piercing the corresponding matter curves, and the superscript $16^{H}$ is used here to denote the origin from $27_{t_{3}}$.
\noindent For the number of $10$'s of $SO(10)$ in the second model, we find one $\overline{10}_{2}$ and $4\times{10^{H}_{-2}}$, and assuming that one pair decouples
(see next section)  we   have                          
\begin{equation}
{10}_{-2}^{H}=
\left\{\begin{array}{ll}{\rm
Rep}&{\rm flux\, units }\\
\ 5_{2}& \;3+n_{\chi}' \\
\ \bar{5}_{-2}& \; 3+n_{\chi}''\\
\end{array}\right.
\end{equation}
Choosing $n_{\chi}=-m_{\chi}=1$, we find 
$ 3\times 10_{-1}$ and  $4\times \bar 5_3$ emerging from  $\Sigma_{16_{t_1}}$, 
$1\times  5_{-3}$ from  $\Sigma_{16_{t_3}}$ and three singlet fields. This implies a three family $SU(5)$ spectrum (supplemented  by the right-handed neutrinos), accommodated in $10+\bar 5+ 1$ representations, and an extra
pair of $\bar 5+ 5$.  Furthermore, imposing $n_{\chi}'=n_{\chi}''=0$ the three $10$'s of $SO(10)$ lead  to three pairs of $5_{-2}+\bar 5_2$. In a final step the breaking of $SU(5)$ is  achieved by turning on hypercharge fluxes, so that
the doublet-triplet spliting mechanism is realised. The spectrum is summarised in Table~\ref{spectrum}. 
 In the following sections we discuss the basic features of the effective theory 
and the implications of the extra  matter and the light boson $Z'$ on the gauge and the  Yukawa sector.

 \subsection{Yukawa couplings of the effective model }
 
 After the $E_6$ breaking, the tree-level superpotential  at the $SO(10)$ level contains the terms
 \ba 
{\cal W}_{1}&\supset& \lambda_i 16_{1} 16_{1}10_{-2}^{H_i}+ \kappa_i  10_{-2}^{H_{i}}10_{-2}\, 1_{4}+\mu_i\theta_{31}\, 1_41_{-4}^{H_i}\,.
 \ea  
The first term provides masses  to fermion fields, while for
 $\langle 1_{4}\rangle \ne 0$,  the second part  generates a massive state of $10_{-2}$ through a linear combination with $10_{-2}^{H_{i}}$.
It  transpires  that at tree-level these are the only mass terms for the various $10$-plets.  Indeed, the couplings 
 $ (\lambda_2'  10_{-2}10_{-2}+ \lambda_3'  10_{-2}^{H_{1}}10_{-2}^{H_2})\times 1_4$,   are not possible due to the $t_i$ charges. 
 They only appear at a non-renormalisable level when a certain number of singlets $1_{t_1-t_3}$ are inserted.
Furthermore, we observe that if
 $\theta_{31}$ acquires a vev $ \langle \theta_{31}\rangle \sim  10^{-1} { M}_{GUT}$, then the two pairs of $1_41_{-4}^{H_i}$ become massive.

Next, let us discuss in brief  possible sources of proton decay.   Under further breaking of $SO(10)$ to  $SU(5)\times U(1)_{\chi}$, the decomposition 
of $27/\overline{27}$ give  $10/\ov{10}$'s and $\bar 5/5$'s.  The relevant term for proton decay   can be $U(1)_{\psi}$-invariant  if a singlet 
is introduced, so that the term ${\cal W}\supset 10_{1,-1}^3 5_{1,3}1_{-4,0}$ is
gauge invariant with respect to  $SU(5)\times U(1)_{\chi}$. 
However, the $t_i$ charges emanating from $SU(3)_{\perp}$ spectral symmetry, do not match. In fact, two additional singlets  $\theta_{31}$  are required
to generate the coupling:
\[{\cal W}\supset 10_{1,-1}10_{1,-1}10_{1,-1}5_{1,3}1_{-4,0} \theta_{31}^2 .
 \]
Therefore, this term is highly suppressed.

Finally, let us briefly discuss the possible contributions to the massless spectrum from the $E_6$ adjoint, i.e. bulk states from 
the decomposition of ${78}$.
As has been previously  shown~\cite{Beasley:2008dc}, in groups of rank 5 or higher not all bulk states are eliminated and therefore the zero mode spectrum 
is expected to contain components of ${78}$. It is possible that some of these states remain at low energies.
Although there are some interesting phenomenological implications of such states~\cite{Callaghan:2013kaa}, 
in the present work we will assume that  they become massive at some high scale and will therefore not be included in our analysis.

\begin{center}
\begin{table}[t]
\centering
\begin{tabular}{|c|cc|}
\hline
Spectrum & $SU(5)$ &$SO(10)$   \\
$3\times{(Q,u^{c},e^{c})}$ & 10 & 16 \\
$3\times{(d^{c},L)}$ & $\ov{5}$ & 16\\
$3\times{\nu^{c}}$ & 1 & $16/\overline{16}$ \\
$3\times{\ov{D},4\times H_{d}}$ & $\ov{5}$ &$ 10/16$ \\
$3\times{D}, 4\times{H_{u}}$ & $5$ &$ 10/\overline{16}$ \\
$S$ & $1$ & $1$   \\
\hline
\end{tabular}
\caption{\small The  spectrum of the effective model and its $SO(10)$ origin used in the RGE analysis. In addition to 	the $H_{u}$ and $H_{d}$ MSSM Higgs pair, three complete $SU(5)$ multiplets in  $5+\bar 5$ 
are assumed to remain in the low energy spectrum. The content of the Table refers to the $N,\eta,\psi$ models.}
\label{spectrum}
\end{table}
\end{center}

\section{RGE analysis for Gauge and Yukawa couplings }

As we have seen, from the decomposition of the $E_6$ representations  there are always additional fields, beyond those of the MSSM spectrum. 
For our RGE analysis we will consider an effective model  that  contains the three families embedded in three 16-plets~$\in SO(10)$,
 where the three right-handed neutrinos decouple at a scale $\sim 10^{14}$ GeV. As shown in the previous section the exact form of the  low energy spectrum and the 
 superpotential depends on specific choices of fluxes, singlet vevs and other parameters, but such an analysis is beyond the scope
 of the present letter. Here, we will focus on  a single case where  additional matter  comprises three complete $SU(5)$  vectorlike
 $5+\bar 5$ multiplets and a  singlet $S$, and the remaining  singlets  $1_4, 1_{-4}$ are assumed to  decouple from the light spectrum.
The MSSM Higgs fields $H_u, H_d$  are accommodated in 5-plets arising from the
$SO(10)$  10-plets  $10_{-2},\overline{10}_{2}$. We suppose that all other components  are removed from the spectrum either by appealing 
to fluxes  or due to a possible doublet-triplet splitting mechanism through couplings  with the bulk states. 
 Under these assumptions,  we have the particle content  presented in Table \ref{spectrum}.

 The computation of the 2-loop RGE's was performed with the use of  the  Mathematica code \texttt{ SARAH-4.10.0}~\cite{Staub:2013tta}. We consider only the Yukawa couplings of the third generation (called here as $Y_{t}$, $Y_{b}$ and $Y_{\tau}$) and for simplicity, we neglect the effects of $U(1)$ kinetic mixing\footnote{An analysis of the effects of $U(1)$ mixing at the 2-loop level is presented in \cite{Cho:2016afr}.}. We take $M_{SUSY}=1$ TeV, $M_{S}=8$ TeV and  a Majorana scale $M_{N}=10^{14}$ GeV, where the heavy right-handed neutrinos decouple from the theory, while all the other extra particles decouple at the scale $M_{S}$. 
\begin{figure}[t!]
\begin{subfigure}{.5\textwidth}
  \centering
  \includegraphics[width=.95\linewidth]{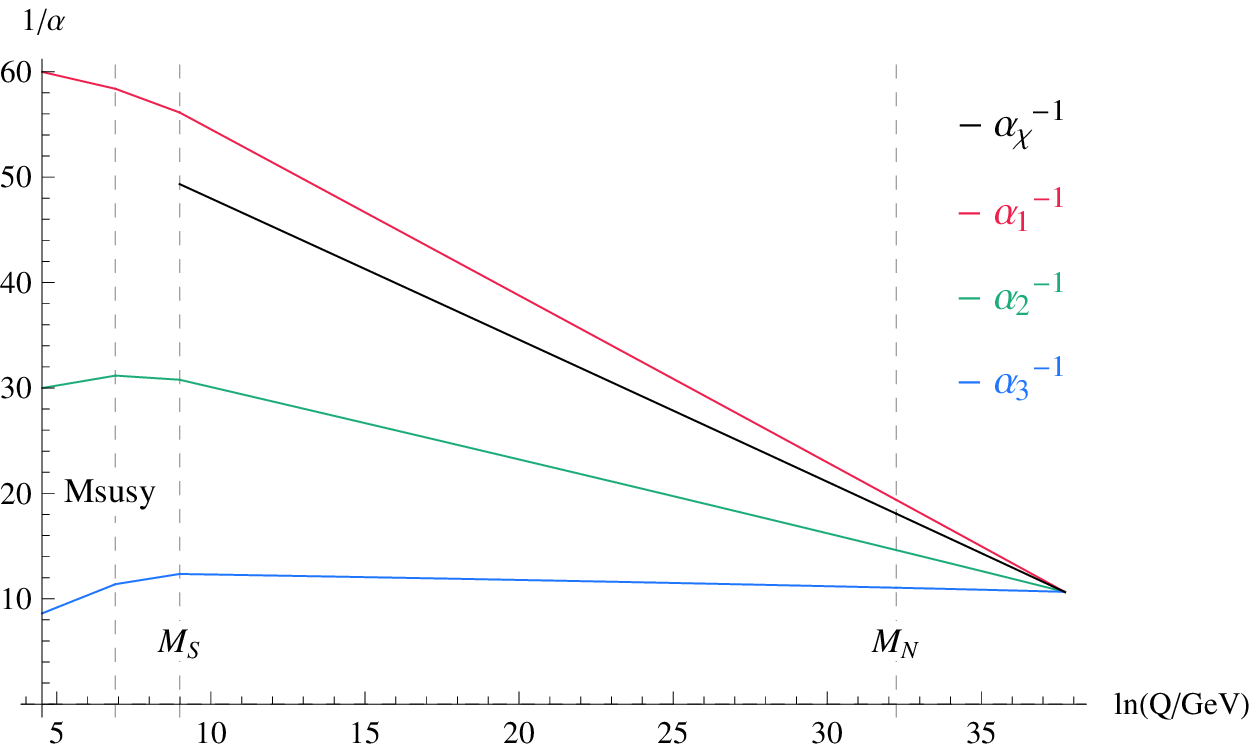}
  \caption{$U(1)_{\chi}$}
  \label{fig:sfig1}
\end{subfigure}%
\begin{subfigure}{.5\textwidth}
  \centering
  \includegraphics[width=.95\linewidth]{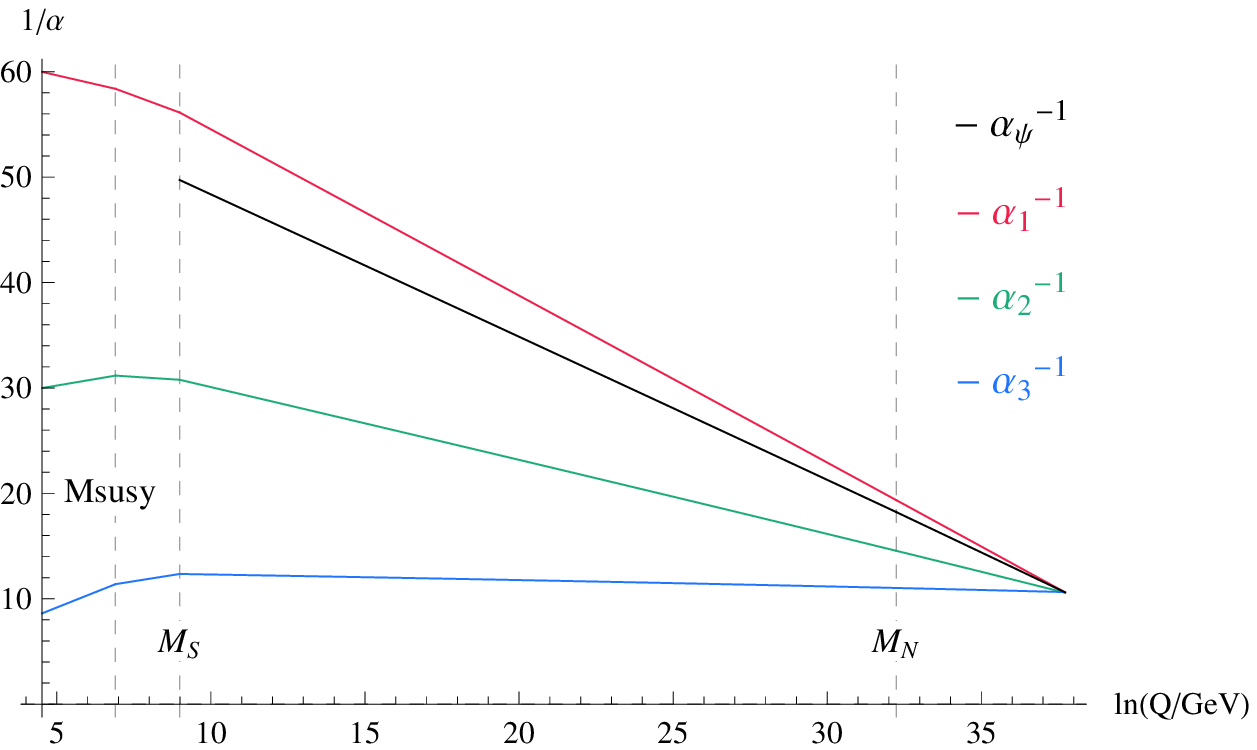}
  \caption{$U(1)_{\psi}$}
  \label{fig:sfig2}
\end{subfigure}
\begin{subfigure}{.5\textwidth}
  \centering
  \includegraphics[width=.95\linewidth]{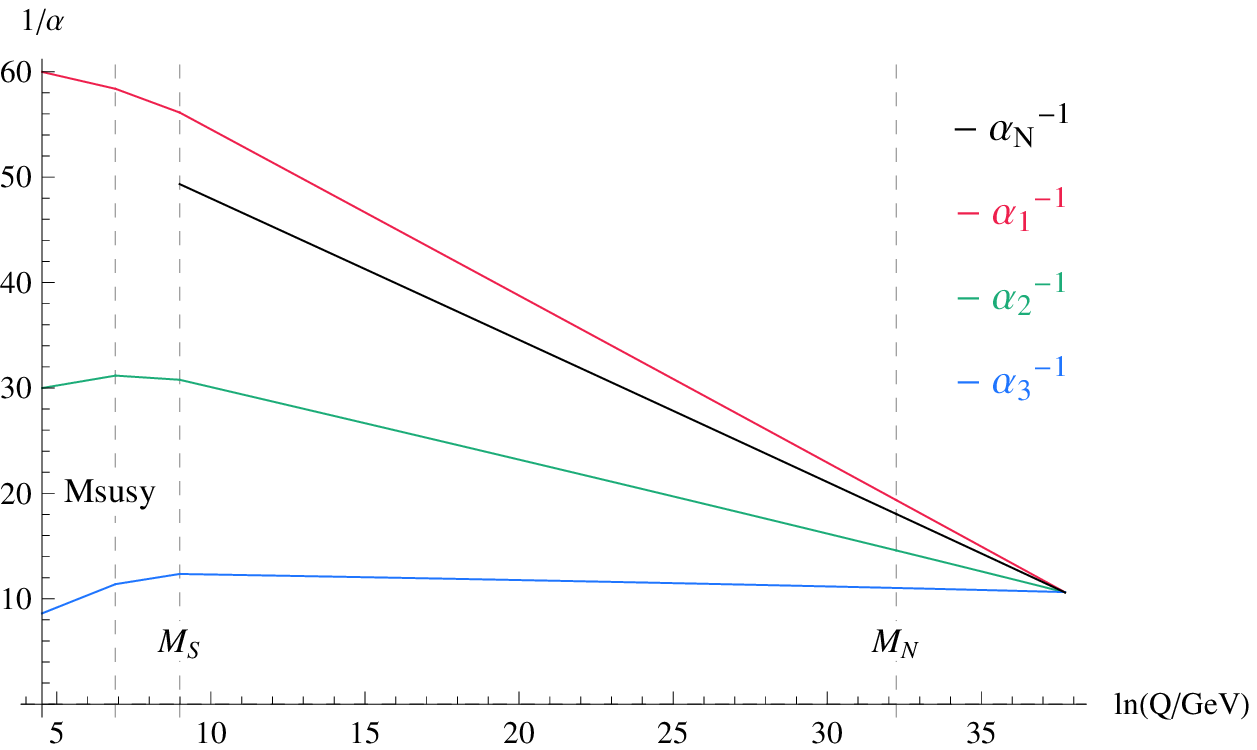}
  \caption{$U(1)_{N}$}
  \label{fig:sfig3}
\end{subfigure}%
\begin{subfigure}{.5\textwidth}
  \centering
  \includegraphics[width=.95\linewidth]{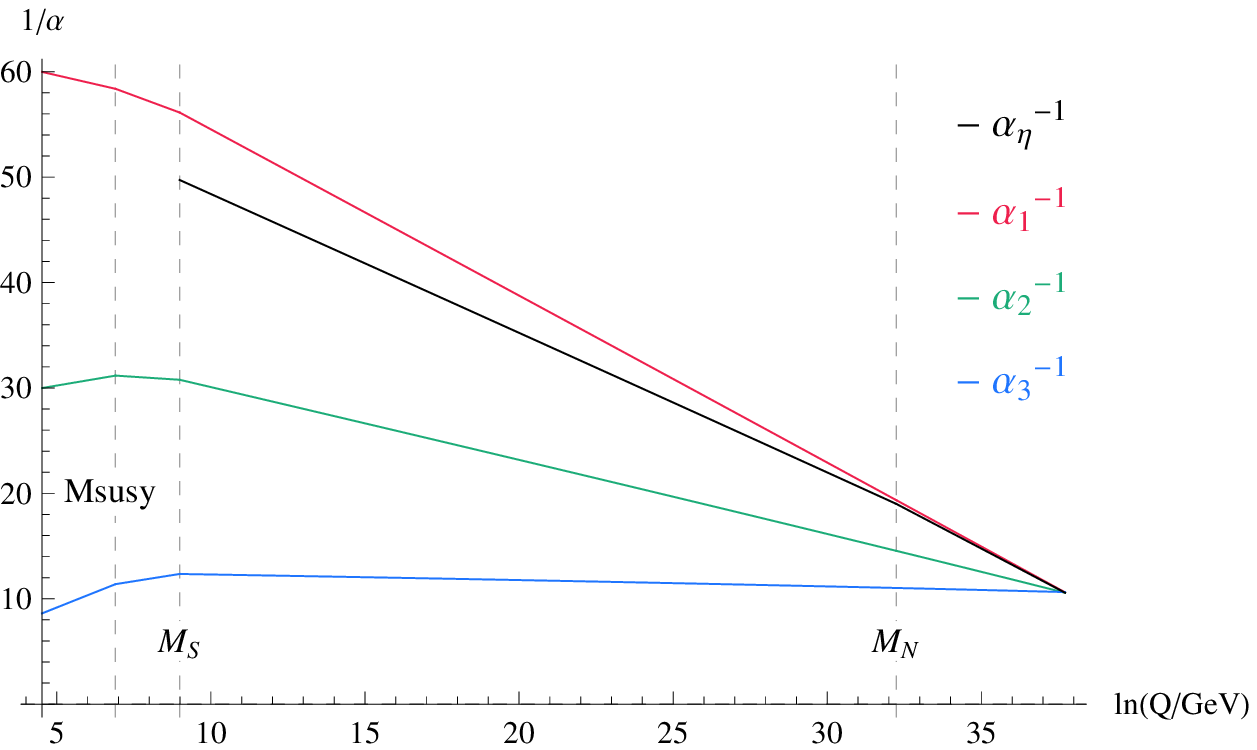}
  \caption{$U(1)_{\eta}$}
  \label{fig:sfig4}
\end{subfigure}
\caption{\small Gauge coupling unification in $E_6$ models. In all cases $M_{GUT}=2.4\times{10^{16}}$  GeV with $g_{U}\simeq{1.09}$. Here $M_{SUSY}=10^{3}$  GeV, $M_{S}=8\times{10^{3}}$  GeV and $M_{N}=10^{14}$ GeV.}
\label{gaugecouplings}
\end{figure}

Using the  mass scales and parameters as described  above, we obtain values of the three SM gauge couplings within the range constrained by the experimental results. 
In Figure~\ref{gaugecouplings} we present their evolution together with the abelian factor corresponding to the $U(1)_{\chi},U(1)_{\psi},U(1)_{N},$ and $U(1)_{\eta}$ models 
respectively. As shown in the figure, the decoupling  of $U(1)$ is assumed at the mass scale $M_S= 8$ TeV. 
 The beta coefficient of the extra $U(1)$ gauge coupling depends on the corresponding charge as follows:
\begin{equation}
b_{\chi}=163/20,\quad b_{\psi}=25/3,\quad b_{N}=163/20,\quad b_{\eta}=227/30.   
\end{equation}

\noindent By assuming unification at $M_{GUT}=2.4\times{10^{16}}$ GeV we obtain the following values for the extra gauge coupling at the scale $M_{S}=8$ TeV : 

\begin{equation}
g_{\chi}(M_{S})\simeq{0.508},\quad g_{\psi}(M_{S})\simeq{0.506},\quad g_{N}(M_{S})\simeq{0.508},\quad g_{\eta}(M_{S})\simeq{0.506}.
\end{equation}

\begin{figure}[t!]
\begin{subfigure}{.5\textwidth}
  \centering
  \includegraphics[width=.95\linewidth]{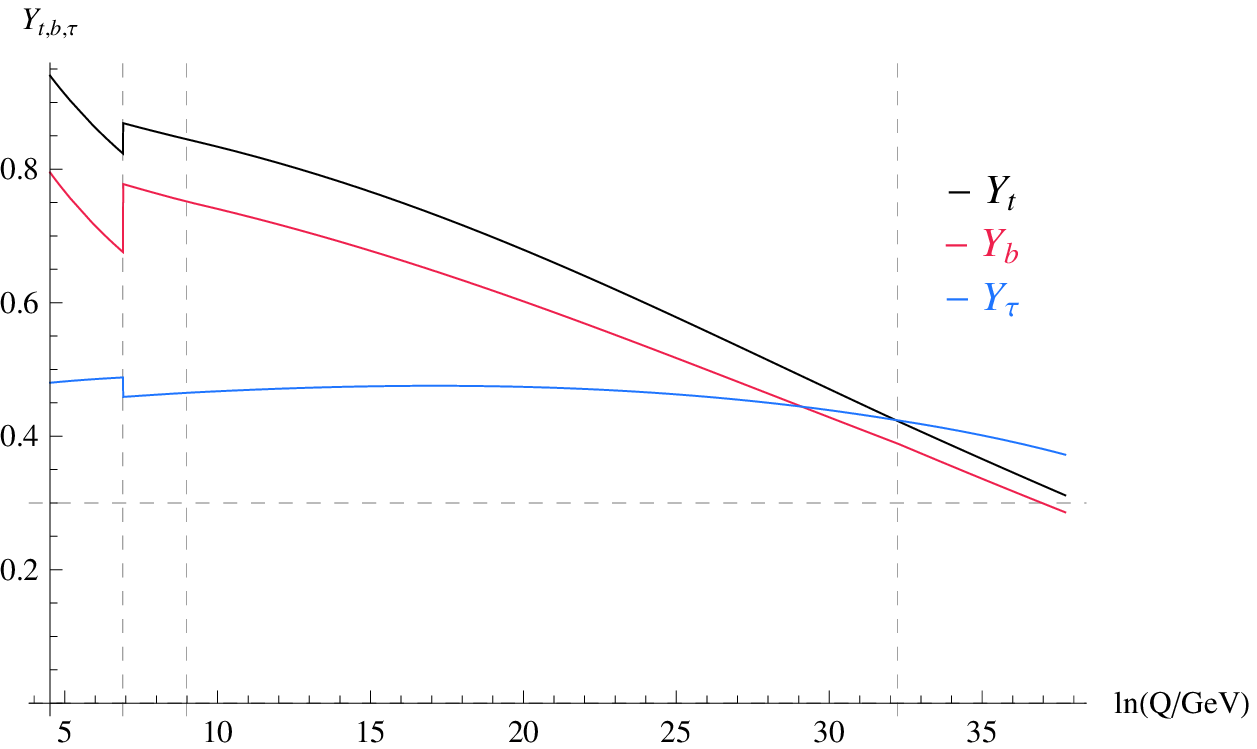}
  \caption{$U(1)_{\chi}$}
  \label{fig:sfig5}
\end{subfigure}%
\begin{subfigure}{.5\textwidth}
  \centering
  \includegraphics[width=.95\linewidth]{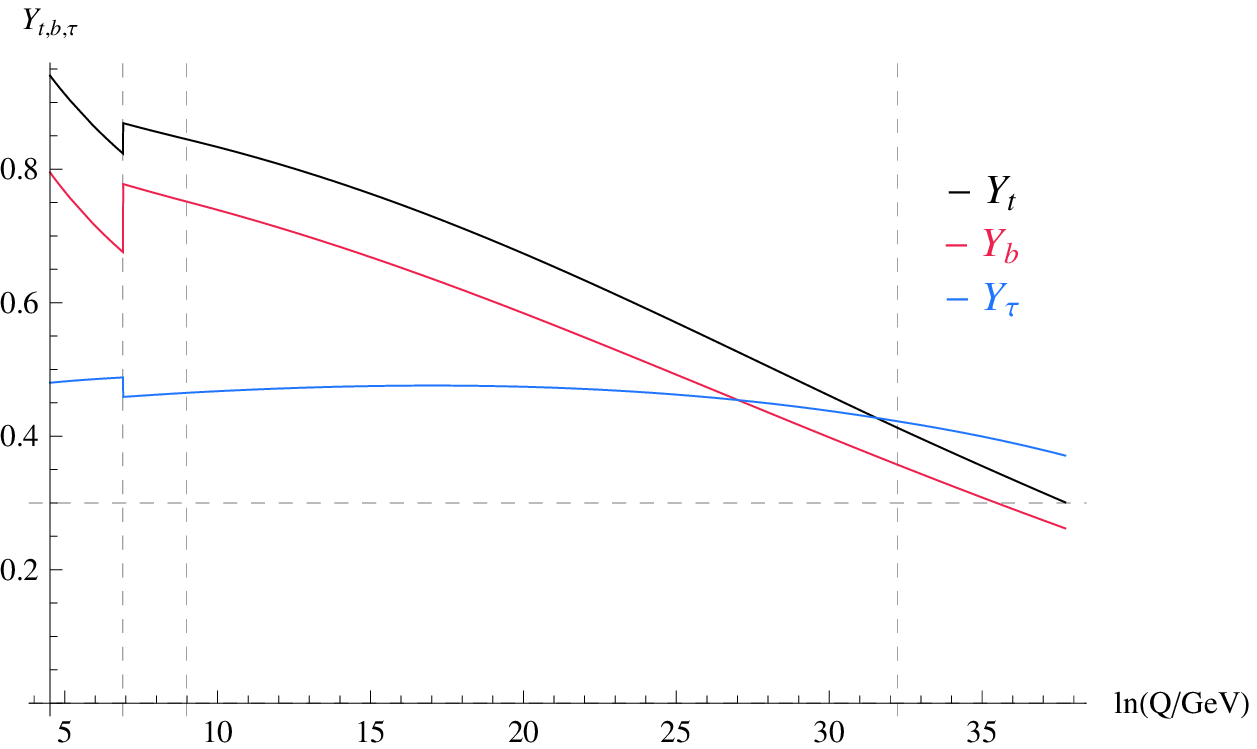}
  \caption{$U(1)_{\psi}$}
  \label{fig:sfig6}
\end{subfigure}
\begin{subfigure}{.5\textwidth}
  \centering
  \includegraphics[width=.95\linewidth]{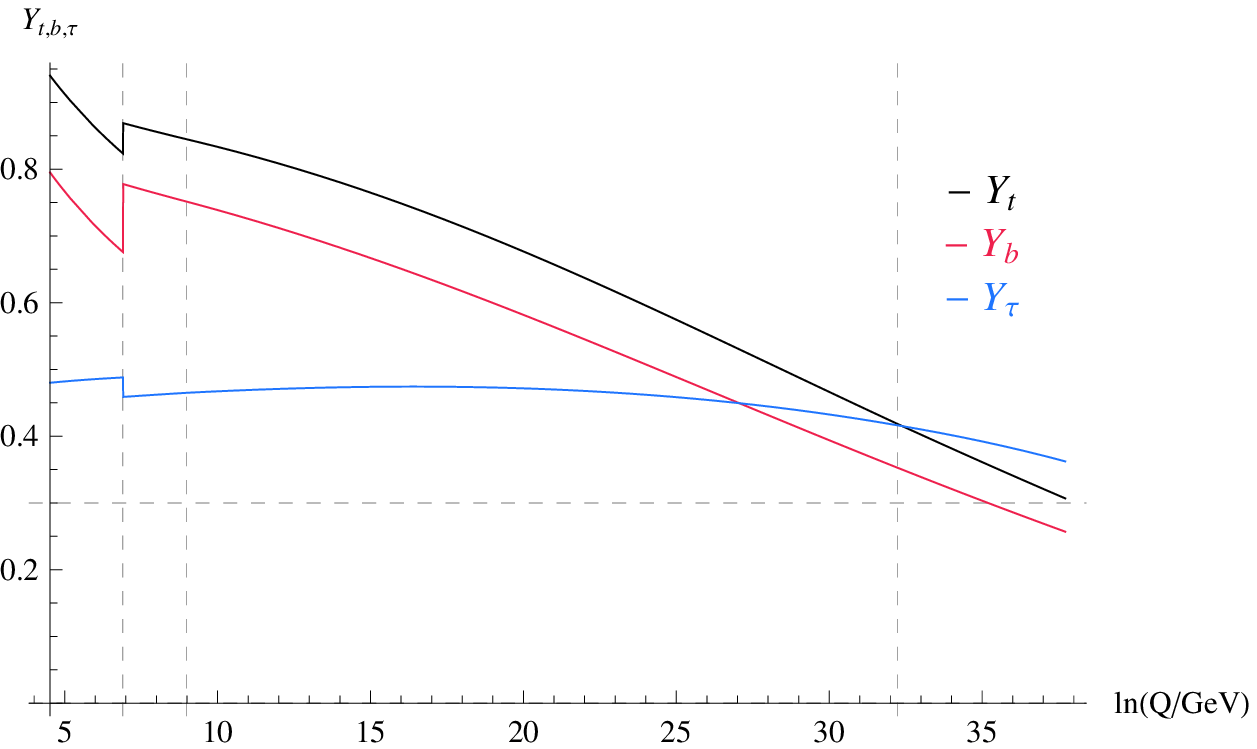}
  \caption{$U(1)_{N}$}
  \label{fig:sfig7}
\end{subfigure}%
\begin{subfigure}{.5\textwidth}
  \centering
  \includegraphics[width=.95\linewidth]{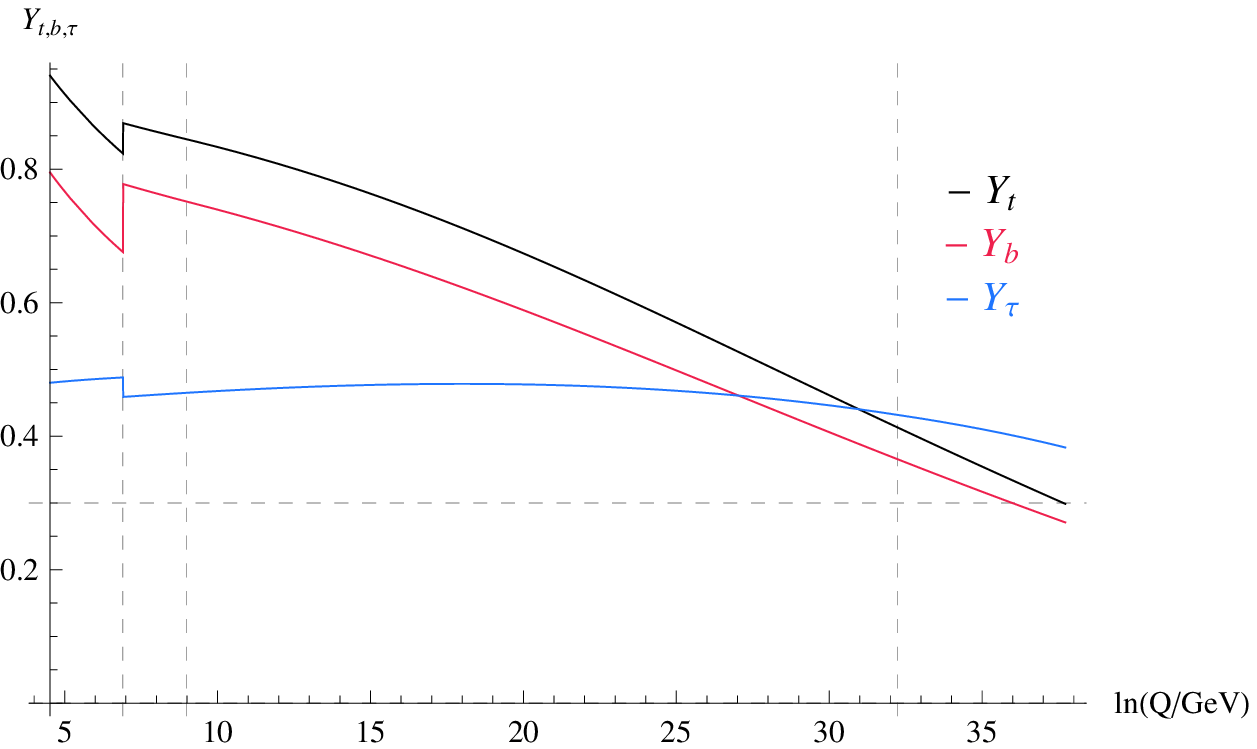}
  \caption{$U(1)_{\eta}$}
  \label{fig:sfig8}
\end{subfigure}
\caption{\small Running of t-b-$\tau$ Yukawa couplings. The horizontal dashed line corresponds to Y=0.3 and is used here for guidance. 
	Here $tan\beta=50$,  $|\mu|=0.5$ TeV and $A_{t}=2.2$ TeV. 
}
\label{yukawacouplings5}
\end{figure}

\begin{figure}[h!]
\begin{subfigure}{.5\textwidth}
  \centering
  \includegraphics[width=.9\linewidth]{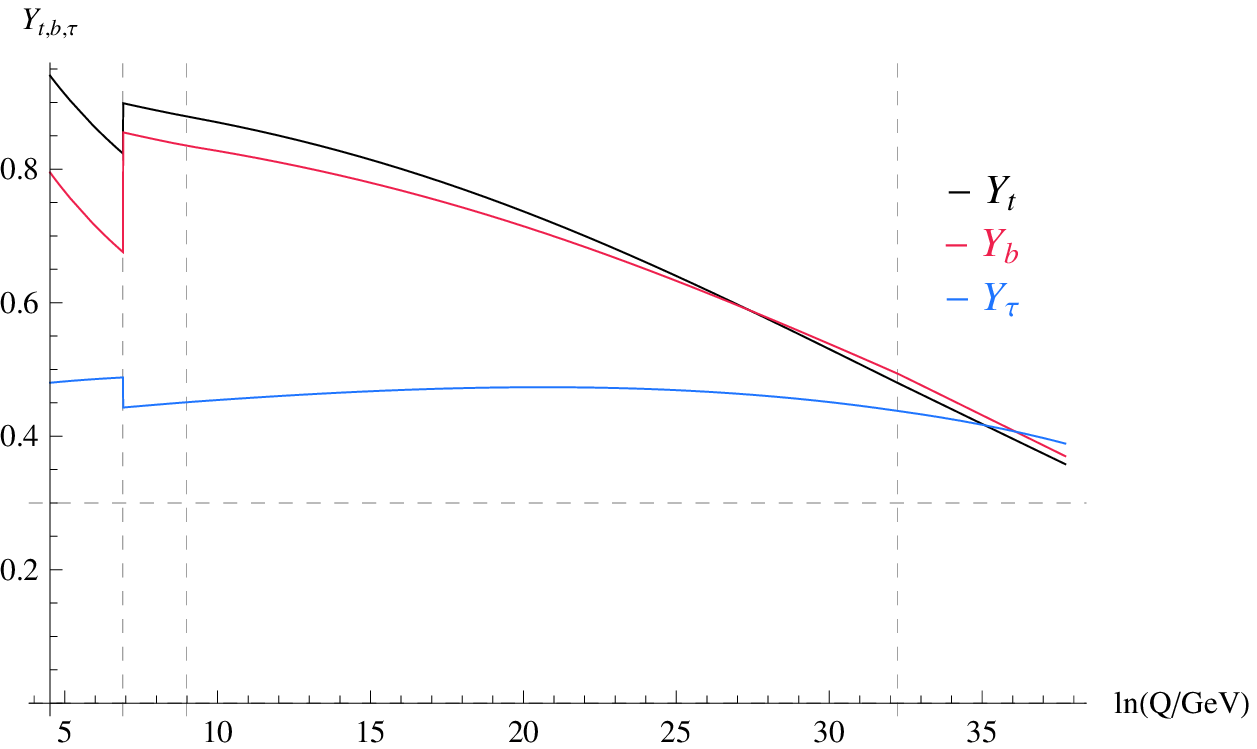}
  \caption{$U(1)_{\chi}$}
  \label{fig:sfig5}
\end{subfigure}%
\begin{subfigure}{.5\textwidth}
  \centering
  \includegraphics[width=.9\linewidth]{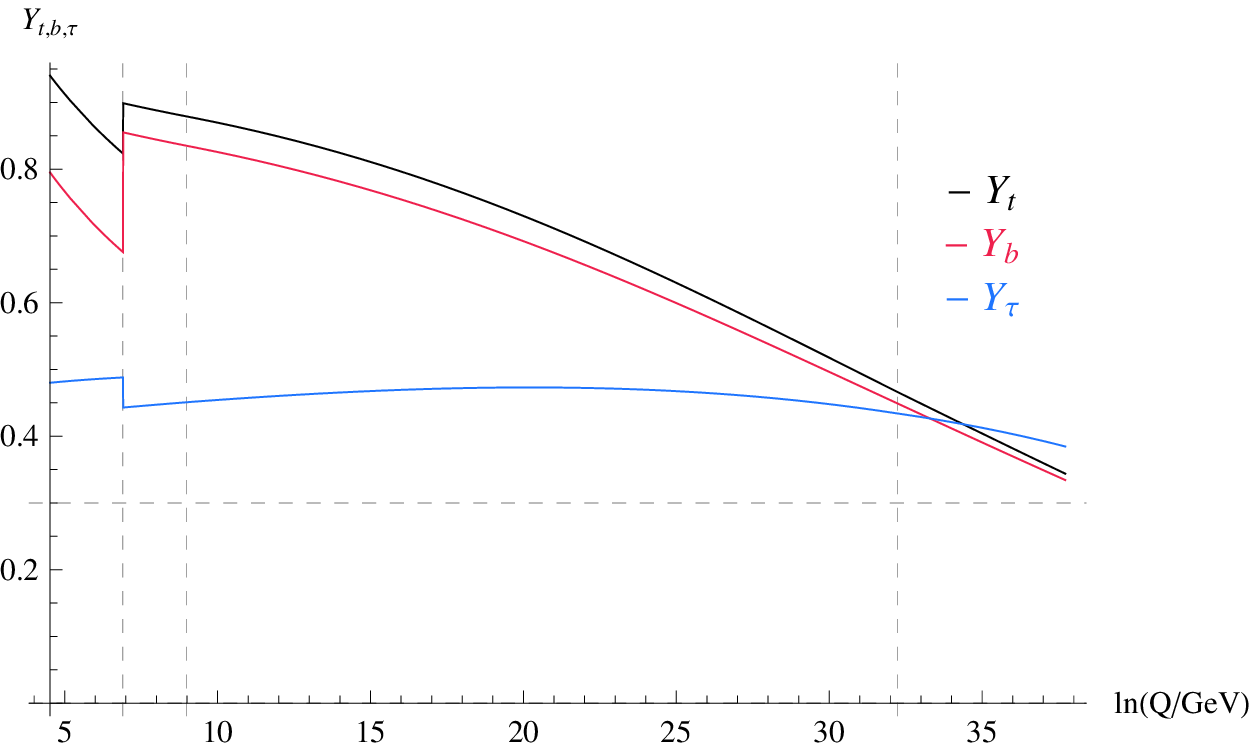}
  \caption{$U(1)_{\psi}$}
  \label{fig:sfig6}
\end{subfigure}
\begin{subfigure}{.5\textwidth}
  \centering
  \includegraphics[width=.9\linewidth]{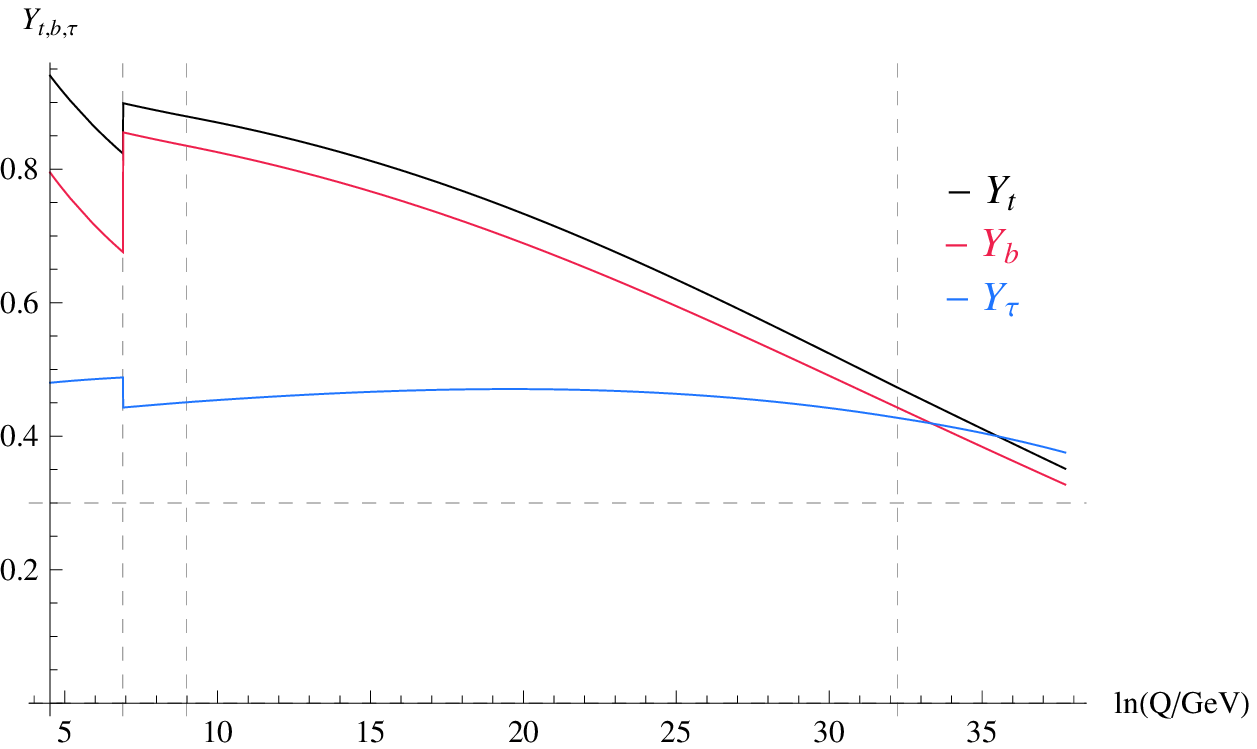}
  \caption{$U(1)_{N}$}
  \label{fig:sfig7}
\end{subfigure}%
\begin{subfigure}{.5\textwidth}
  \centering
  \includegraphics[width=.9\linewidth]{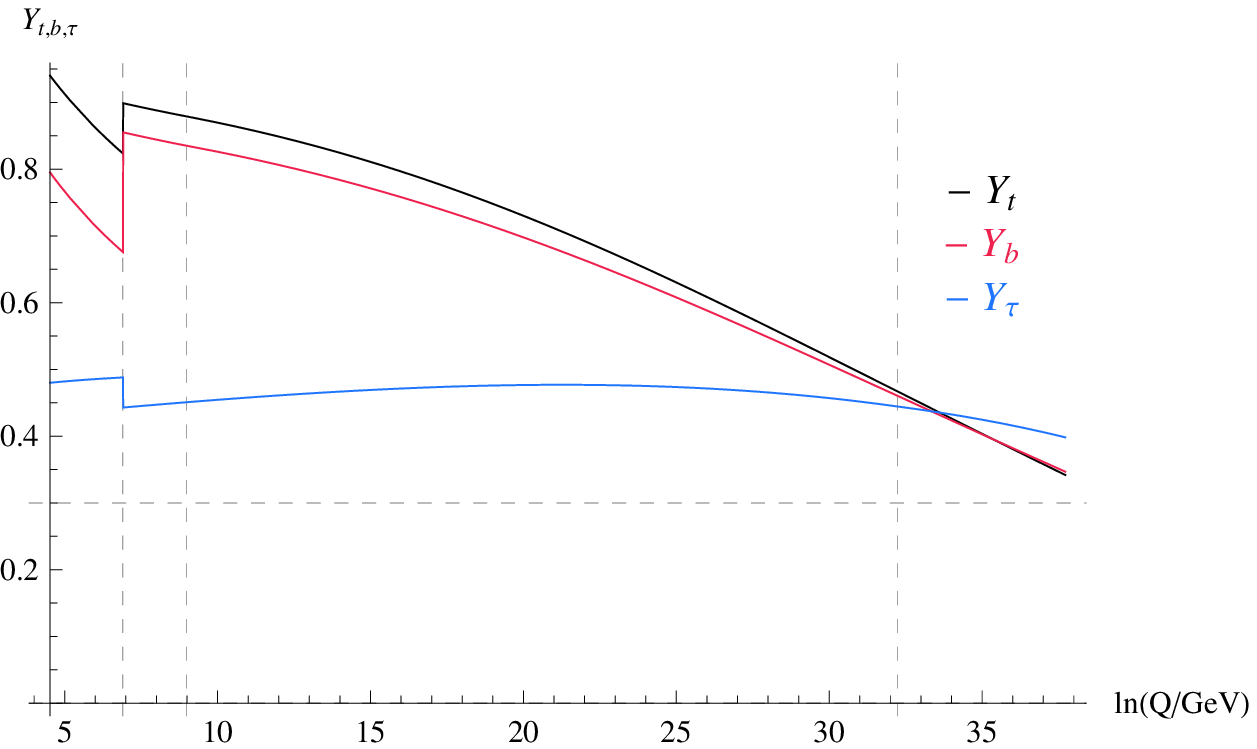}
  \caption{$U(1)_{\eta}$}
  \label{fig:sfig8}
\end{subfigure}
\caption{\small Running of t-b-$\tau$ Yukawa couplings. The horizontal dashed line corresponds to Y=0.3 and is used here for guidance. Here $tan\beta=50$,  $|\mu|=0.8$ TeV and $A_{t}=2.2$ TeV. 
}
\label{yukawacouplings8}
\end{figure}

Next we proceed with the Yukawa sector. 
In Figures~\ref{yukawacouplings5}  and~\ref{yukawacouplings8}  we present the evolution of the third generation Yukawa couplings for $\tan\beta=50$.
Figure~\ref{yukawacouplings5} corresponds to  $|\mu|=0.5$ TeV and Figure ~\ref{yukawacouplings8} to $|\mu|=0.8$ TeV. In both cases,  the masses of the sfermions were taken in the range of $2-3$ TeV and  the trilinear parameter  $A_{t}=2.2$ TeV.  We observe that, in contrast to the minimal spectrum, 
in the presence of additional vectorlike matter,  a moderate value of the top Yukawa coupling at the GUT-scale  can  reproduce the top mass at the electroweak scale.  Furthermore, comparing  Figures~\ref{yukawacouplings5}  and~\ref{yukawacouplings8}, we see that an increment of  the SUSY threshold corrections and the value of $|\mu|$,  implies larger GUT values of the Yukawa couplings. Some  representative values for the same SUSY parameters but 
two different values of  $\mu$ are presented in Tables~\ref{values5} and \ref{values8}. Our findings show that  the  results are the same for $\chi$ 
and $N$ models. For a discussion of sparticle spectroscopy with t-b-$\tau$ Yukawa unification see~\cite{Ajaib:2014ana} and references therein.

We close this section with a few observations.   First, we notice that 
raising  the  scale $M_{S}$ by a few TeV increases slightly the value of the Yukawa couplings. At the same time we get a lower value of the gauge coupling   $g_{U}$ at $ M_{GUT}$.
\begin{center}
\begin{table}[h]
\centering
\begin{tabular}{|c||ccccc|}
\hline
$E_6$ model & $Y_{t}$ & $Y_{b}$ & $Y_{\tau}$ & $Y_{H}$ & $Y_{D}$\\
\hline
$U(1)_{\chi}$ & 0.305 & 0.257 & 0.361 & 0.336 & 0.306\\

$U(1)_{\psi}$ & 0.300 & 0.262 & 0.370 & 0.330 & 0.300\\

$U(1)_{N}$ & 0.305 & 0.257 & 0.361 & 0.336 & 0.306\\

$U(1)_{\eta}$ & 0.297 & 0.270 & 0.380 & 0.345 & 0.324\\
\hline
\end{tabular}
\caption{\small Numerical values of the Yukawa couplings at $M_{GUT}$ for $\tan\beta=50$ and $|\mu|=0.5$ TeV.  
	The last two columns refer to the Yukawa couplings of the  vectorlike pairs.  }
\label{values5}
\end{table}
\end{center}

The $Z'$ boson mass for the various models discussed above are as follows:
 \be  
 M_{Z_{\psi}}\approx 4.67 \; {\rm TeV},\;\;M_{Z_{N}}\approx 4.54 \; {\rm TeV},\;
M_{Z_{\eta}} \approx  3.70 \; {\rm TeV}.
\ee 

\noindent In all cases, the predicted mass of $Z'$  lies just above the current experimental bounds 
given by~\cite{Khachatryan:2017wny,Aaboud:2017yvp,ATLAS:2017wce}
	\[M^{exp}_{Z'}\,>\,3.4-4.1  \,{\rm TeV}~.\]

\begin{center}
\begin{table}[h!]
\centering
\begin{tabular}{|c||ccccc|}
\hline
$E_6$ model & $Y_{t}$ & $Y_{b}$ & $Y_{\tau}$ & $Y_{H}$ & $Y_{D}$ \\
\hline
$U(1)_{\chi}$ & 0.350 & 0.326 & 0.374  & 0.361 & 0.350\\

$U(1)_{\psi}$ & 0.342 & 0.333 & 0.383  & 0.372 & 0.358\\

$U(1)_{N}$ & 0.350 & 0.326 & 0.374 & 0.361 & 0.350\\

$U(1)_{\eta}$ & 0.340 & 0.345 & 0.396 & 0.372 & 0.371\\
\hline
\end{tabular}
\caption{\small Numerical values of the Yukawa couplings at the GUT scale for $\tan\beta=50$ and $|\mu|=0.8$ TeV.     
 The last two columns refer to the Yukawa couplings of the third family vectorlike pairs. }
\label{values8}
\end{table}
\end{center}

Next we discuss the extra doublet and vectorlike color triplet fields. As an example, following~\cite{Hebbar:2016gab},
we assume that the Yukawa couplings, $Y_{H}$ and $Y_D$, of one pair $H_{u}+H_{d}$ and one pair $D+\bar D$, unify  asymptotically 
with the Yukawa couplings of the third generation at the GUT scale. 
The values of these couplings at the GUT scale are also presented in Tables \ref{values5} and \ref{values8}. Using the RGE's we 
predict the value at the scale $M_S$.  We find that the masses of  $D+\bar{D}$ and the extra $H_{u}+H_{d}$ doublets are:
\begin{align}\label{exoticsbound}
m_{D} & \geq{5.92}\quad \text{TeV},\\
m_{H}& \geq{3.44}\quad \text{TeV}.
\end{align}

Finally, in our analysis we have found that  in the presence of extra vectorlike pairs and singlet fields at a few TeV scale,
the third generation fermion masses and in particular the top-mass can be correctly reproduced with moderate values of the Yukawa 
couplings at the GUT scale. As we will show, this is  in agreement with the predictions from F-theory computations.

\subsection{Yukawa Couplings in F-Theory}

In F-theory, the Yukawa couplings are realised when three Riemann surfaces accommodating matter fields  intersect at a single point 
on the GUT surface, $S$. Given the specific geometry of the compact space, we can solve the appropriate equations of motion and determine  
the profile of the wavefunctions of the states involved. The Yukawa couplings are then obtained by computing the integral of the
 overlapping wavefunctions at the triple intersections. The final result of the computation depends on local flux densities permeating 
 the matter curves.  In the present work, we consider an $E_8$ point of enhancement and follow the procedures 
 described  in a series of papers~\cite{Cecotti:2010bp}-\cite{Marchesano:2015dfa}. We should note that the
 flux units considered in Section 2 are integer valued as they arise from the Dirac quantisation 
 \begin{equation}
 \frac{1}{2\pi}\int_{\Sigma\subset{S}}F=n_f\,,
 \end{equation} 
 where $n_f$ is an integer, $\Sigma$ denotes a matter curve (two-cycle in the divisor $S$), and $F$ is the gauge field strength tensor,  i.e., the flux.
 In the same section we also described how the flux units piercing different matter curves $\Sigma$ determine the chiral states which are globally present in a
 given model.  However, while the flux units in Section 2 define the full spectrum of the model, the study of the trilinear couplings involve the calculation of 
 the wavefunctions and their overlaps on a local, approximately flat patch around a point of intersection. In this local approach it is the local values of 
 flux -and not the global quantisation constraints- that matter. The local fluxes determine the chiral states at the local point. Besides those,  there can be 
 additional chiral fermions localised in other regions of the matter curve, with the total chirality determined by the integral of the magnetic flux along the 
 matter curve. The relation between local and global fluxes is not a clear issue since it requires a complete knowledge of the geometry of the matter curve.  
  A more sophisticated \emph{local vs. global} analysis is given in \cite{Palti:2012aa}.
 In our present approach, we will consider ranges of flux densities corresponding to a wide range of integer values encompassing 
 also those flux parameters used in section 2.
 
Following the formulation of \cite{Marchesano:2015dfa} (see also~\cite{CrispimRomao:2016tww}) we deal with two types of flux 
density parameters.  The first type is parametrised by the flux density numbers $M_{i}$, $N_{i}$ where $i=1,2$,
and descend from a worldvolume flux which is necessary to  induce chirality on the matter curves accommodating the 
 $10$-plets, $\bar 5$-plets and $5$-plets  of $SU(5)_{GUT}$. The second type parametrised by $N_{Y}$ and $\tilde{N}_{Y}$, is related 
 to the hypercharge flux which breaks the $SU(5)$ symmetry to the Standard Model and in addition generates the observed chirality 
 of the fermion  families. 

\begin{figure}[h!]
  \centering
  \includegraphics[width=.8\linewidth]{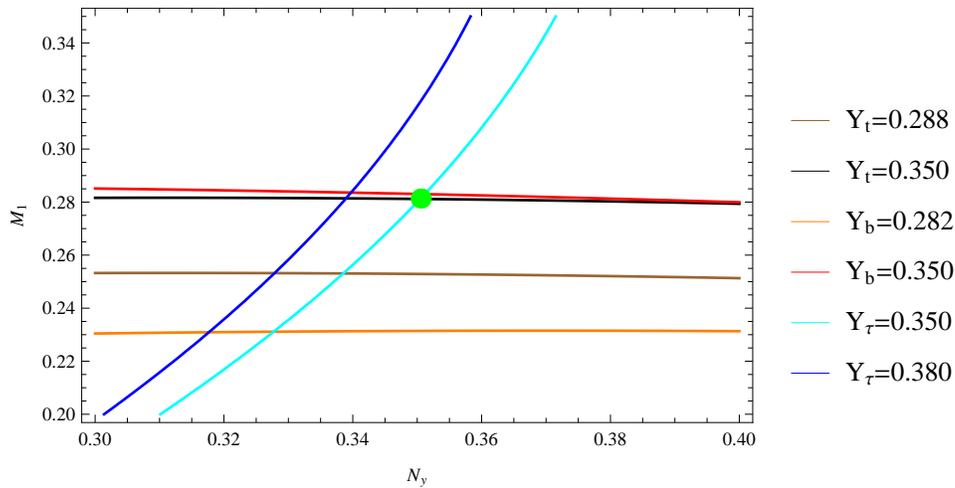}
  \caption{\small Values of the Yukawa couplings from the $E_8$ point in
  	 F-theory without imposing any constraint on the flux parameters. 
  	 Green point corresponds to $Y_{t}\approx{Y_{b}}\approx{Y_{\tau}}=0.35$.}
  \label{fsol1}
\end{figure}

In Figure~\ref{fsol1} we plot  the bottom, tau and top Yukawa coupling at the 
 local flux-density parameter space $M_{1}$ and $N_{Y}$. For the remaining flux density parameters involved in the
 computation we consider the values $N_{1}=0.187, M_{2}=1.23, N_{2}=0.701, \tilde{N}_{Y}=0.09$. 
  For a reasonable range of the $M_{1}$ and $N_{Y}$ parameters, the values of  $Y_{t,b,\tau}$ lie approximately
 between $0.3$ and  $0.4$. There is a single  $(M_{1}, N_{Y})$ point 
 (shown with green color bullet in Figure~\ref{fsol1}) where all Yukawa 
 couplings of the third generation attain the same value $Y_{t,b,\tau}=0.35$.

Before closing this section, we make a few comments regarding the issues emerging from supersymmetry breaking,
such as soft masses and flavour changing neutral currents (FCNC). The structure of the SUSY breaking soft terms have been studied
for a large class of string and flux compactifications with a MSSM-like spectrum~\cite{Brignole:1993dj}-\cite{Choi:2008hn}. In many cases the presence of non-diagonal flavor dependent SUSY-breaking soft terms are generically induced. The presence of such terms can lead to dangerous FCNC effects which can create tension with other phenomenological predictions of the low energy theory. In the case of F-theory generalisations, SUSY breaking soft terms and its phenomenological implications have been extensively discussed 
in the past~\cite{Aparicio:2008wh}-\cite{Camara:2014tba}, \cite{Camara:2011nj}. Especially in~\cite{Camara:2013fta}, \cite{Camara:2014tba}, it is shown how SUSY breaking soft terms for fields 
on matter curves are generated from closed string fluxes, applying the results on F-theory local models and including  contributions from magnetic fluxes. In the special 
case of non-constant fluxes flavor dependent soft terms arise which must lie in the multi-TeV range in order to avoid FCNC effects. However, the results strongly depend on the internal geometry, the background fluxes and there is considerable uncertainty from model dependent factors. On the other hand these flavor violating effects may be suppressed if the close string fluxes vary slowly over $S$.

Gravity mediated SUSY breaking is also a possible source of FCNC after integrating out heavy modes. In F-theory local models this scenario has been discussed in \cite{Camara:2011nj} where it is shown that off-diagonal  terms are not induced due to the presence of geometric $U(1)$ symmetries, while a full 
study  of FCNC requires the study of the difference $m^{2}_{22}-m_{11}^{2}$ of the soft scalar masses $m_{ij}$. We expect that this will be suppressed for a wide range 
of the parameter space while a detailed computation is beyond  the scope of this letter.

\section{Conclusions}

In this work, we have presented  effective field theory  models embedded in $E_6$ with an extra neutral  
gauge boson ($Z'$) and  additional vectorlike  fields in the low energy spectrum.  The extra matter fields 
(beyond the MSSM spectrum), assumed to  remain   at the TeV region  include triplets and doublets comprising 
three complete $5+\bar 5$-plets of  $SU(5)$, as well as  neutral singlets. It is shown that this spectrum can 
be embedded naturally in an F-theory scenario where abelian fluxes are used to break the $E_6$ symmetry  to $SU(5)$. 
\noindent
Using renormalisation group analysis at two-loop level, we explore  the implications of this spectrum on the running 
of the gauge and Yukawa couplings. We perform this analysis by  assuming a  $Z'$ boson mass compatible with the 
LHC bounds and  masses of the  extra fields $\sim$~10~TeV, and we take into account  threshold corrections of SUSY 
particles and a right-handed neutrino scale  $10^{14}$ GeV. We find that moderate values at the GUT scale of the third generation
Yukawa coulings in the range $Y_{t,b,\tau}\sim 0.3-0.4$  and $\tan\beta\sim 50$ can successfully reproduce  their low energy masses. 
Finally, based on previous detailed work on Yukawa couplings in  F-theory~\cite{Cecotti:2010bp}-\cite{Marchesano:2015dfa},
we compute the third generation Yukawa couplings generated by a configuration of intersecting seven-branes with the GUT divisor.
We assume a configuration  with a single  $E_8$ point of enhancement and compute the relevant integral taking into account  
non-trivial fluxes associated with the symmetry breaking.  We express the results  in terms of the local flux densities  
and find that  their values are in the same range with those found by the  renormalisation group analysis using as inputs
the known low energy masses of the charged fermions of the third family. We also find points in the 
parameter space of the flux densities where $t-b-\tau$ Yukawa couplings attain a common value.

\vspace{.5cm}
{\bf Acknowledgements}. {
	G.K.L. would like to thank the Physics and Astronomy Department and Bartol Research
	Institute of the University of Delaware, and LPTHE of  UPMC in Paris,
	  for kind hospitality where part of this work has been done.
	 Q.S. is supported in part by the DOE grant DE-SC0013880.}

\end{document}